# Production of Over-dense Plasmas by Launching 2.45GHz Electron Cyclotron Waves in a Helical Device

R. Ikeda[a], M. Takeuchi[a], T. Ito[a], K. Toi[b], C. Suzuki[b], G. Matsunaga[c], S. Okamura[b], and CHS Group[b]

[a] *Department of Energy Engineering and Science, Nagoya University, Nagoya, Japan*
[b] *National Institute for Fusion Science, Toki, Japan*
[c] *Japan Atomic Energy Research Institute, Naka, Japan*

For production of low temperature plasmas with low collisionality, 2.45GHz microwave power up to 20kW is injected perpendicularly to the toroidal field at very low toroidal field $B_t$<0.1T. In the case $B_t$=0.0613T, the maximum electron density reaches three times of the O-mode cutoff density and the measured power deposition is concentrated in the plasma core region beyond the Left-hand cutoff layer. It clearly suggests that this over-dense plasma is produced and heated by electron Bernstein waves converted from the launched X-mode in the peripheral region with steep density gradient.

## 1. Introduction

A transport simulation based on "dimensional similarity" idea using a low temperature and density helical plasma is attempted, where principal dimensionless plasma parameters, such as the normalized electron-ion collision frequency, beta, scale length of temperature and of density and others are the same, except the normalized Larmor radius is different[1-4]. Here, we aim at generating this low temperature and density plasma at very low toroidal field ($B_t$<0.1T) with 2.45 GHz electron cyclotron wave (ECW). In our experiments, the plasma has relatively low collisionality, and the normalized collision frequency already reaches the higher end of 1/ν regime. However, the value of beta has only about 1/5 compared with the value achieved in high magnetic field experiments.

In order to extend these dimensionless parameters, it is important to find the optimum condition so that incidence power should be efficiently absorbed to low temperature plasma. We pay attention to not only ECW but also Electronic Bernstein Wave (EBW)[4]. It is known that EBW can heat over-dense plasma without suffering from the cutoff density, and high absorption is expected even in low temperature plasma. However, since EBW is electrostatic wave and cannot transmit in vacuum, the mode conversion scenario such as X-B or O-X-B is required[5-,7]. In this paper we present the experimental results on the production of over-dense plasma where the mode conversion into EBW is expected.

## 2. Experimental Device and diagnostic tools

Plasma production and heating using 2.45GHz microwaves up to 20kW is carried out at $B_t<0.1$T in the Compact Helical System (CHS)[8], which is a heliotron/torsatron ($R$=1.0m, $a$=0.2m). This plasma is produced with hydrogen gas in 7.0x10$^{-5}$Torr. The launching methods of microwave are the following ways. One is the perpendicular injection of TE$_{10}$-mode from the outboard side port in the horizontally elongated section, of which mode has the electric field perpendicular to the toroidal field, that is, this launching method expects the perpendicular injection of X-mode (Fig.1(a)). The other hand is the oblique injection of TE$_{01}$-mode, of which mode has the electric field parallel to the toroidal field. This method is the oblique injection of O-mode.

A triple-Langmuir probe is employed to simultaneously measure electron density ($n_e$), electron temperature ($T_e$) and space potential. This probe is installed on CHS as shown in Fig.1(b). The central line averaged electron density ($n_e^{2mm}$) is measured by 2mm microwave interferometer. The local electron density measured by the probe is calibrated by $n_e^{2mm}$. In addition, H$\alpha$ emission and total radiation power loss are monitored by a visible spectrometer and a bolometer, respectively.

## 3. Experimental Results

### 3.1 Plasma production and heating by perpendicular injection of X-mode

First, the results of perpendicular injection of X-mode are following. Although the polarization of wave is rather difficult due to strongly twisted magnetic field line in a helical plasma, we tried to change the polarization of launching waves through the change of TE$_{10}$ or TE$_{01}$ mode with or without a twisted waveguide. However, no obvious differences in plasma heating have been observed. Presumably, the polarization of launching waves is opaque because the wavelength of the wave in the vacuum is fairly large to be about half of the plasma minor radius. In the present condition, mixture of O-mode and X-mode will be injected.

The contour plots of the magnetic field strength in the horizontally-elongated section in which an injection port is attached are shown in Fig.2 for two different field strength of $B_t$=0.0788T or $B_t$=0.0613T at the magnetic axis position of the vacuum field $R=R_{ax}$, where $R_{ax}$=97.4cm. As seen from these figures, in the case of $B_t$=0.0788T, the fundamental electron cyclotron resonance (ECR) layer is placed near the plasma center. In the case of $B_t$=0.0613T, the fundamental ECR layer is outside the last closed flux surface (LCFS). The second harmonic cyclotron layer resides in the outer plasma region for both cases. Therefore, the absorption by the fundamental ECR would not be expected in the case of $B_t$=0.0613T where waves are injected perpendicularly to the magnetic field line. It should be noted that the second harmonic cyclotron damping in the peripheral plasma region might be possible in the latter case. The results in both cases are described in this section.

A typical time evolution of the microwaves-produced plasma at $B_t$=0.0788T is shown in Figs.3(a). The line averaged electron density $n_e^{2mm}$ exceeds the O-cutoff density ($n_{c,o}$=7.4x10$^{16}$m$^{-3}$) for a short time interval from 50ms to 80ms. As shown in Fig.3(a), injected microwave power ($P_{in}$) is stepped down at $t$=150ms. The radial profiles of $T_e$ and $n_e$ at 150ms are shown in Fig.3(b). The absorption power density ($P_{abs}$) is derived using the following equation

$$P_{abs}(\rho) \cong \frac{3}{2}\left( \left.\frac{\partial n_e T_e(\rho)}{\partial t}\right|_{t=t_0-0} - \left.\frac{\partial n_e T_e(\rho)}{\partial t}\right|_{t=t_0+0} \right) \quad (1),$$

across the power step at $t=t_o$. As seen from the profile of $P_{abs}$ in Fig.3(b), most of microwave power is deposited in the higher density region inside the Right hand cutoff (R-cutoff) layer, but nearby the fundamental ECR layer. Moreover, $n_e$ at $t$~150ms is blow the cutoff density in a whole plasma region. Here, we try to check the possibility of EBW heating converted through X-B conversion process. The maximum mode conversion efficiency is estimated by using the following simple relation,

$$C_{\max} = 4\exp(-\pi\eta)(1-\exp(-\pi\eta))$$

$$\eta = \frac{\Omega_e L_n}{c} \frac{\kappa}{\sqrt{\kappa^2 + 2(L_n/L_B)}} \left[ \frac{\sqrt{1+\kappa^2}-1}{\kappa^2 + (L_n/L_B)\sqrt{1+\kappa^2}} \right]^{\frac{1}{2}} \quad (2)$$

,where $\kappa=\omega_e/\Omega_e$[9]. Here, $\omega_e$, $\Omega_e$, $L_n$, and $L_B$ are the electron plasma frequency, electron cyclotron frequency, the scale length of density gradient, and the scale length of magnetic field at UHR layer, respectively. The parameter $\eta$ stands for the optical thickness of the evanescent region. If $\eta$ is less than unity, X-mode can penetrate into the evanescent region. In this case, the values of $\eta$ and $C_{max}$ are respectively about 1.1 and 0.1, so that the possibility of X-B conversion should be very low. Since $n_e$ is lower than $n_{c,o}$, it is concluded that the absorption is realized by the fundamental ECR damping of a portion of O-mode component rather than EBW. Total deposited power evaluated from Fig.3(b) is about 0.12kW across the power step-down ($\Delta P_{in}$~5kW), and corresponds to be very low absorption rate (~ 2.5% ) of the injected power.

Next, we investigated the plasma heating in the latter case of $B_t$=0.0613T. In the shot shown in Fig.4(a), $n_e^{2mm}$ is more than twice $n_{c,o}$ throughout the discharge. The injection power $P_{in}$ is stepped down at $t$=185ms. The radial profiles of $T_e$ and $n_e$ at 185ms are shown in Fig.4(b). As seen from Fig.4(b), $n_e$ exceeds $n_{c,o}$ over the almost whole plasma region ( $\rho \leq 0.9$). The microwave power is absorbed clearly inside the R-cutoff layer and the upper hybrid resonance (UHR) layer placed at the edge region of $\rho$~0.9. This suggests that fast X-mode - slow X-mode - EBW conversion takes place in the edge region[10]. In this process, fast X-mode will tunnel through the evanescent region between the R-cutoff layer and the UHR layer, and then be converted into slow X-mode at UHR layer. Moreover, the slow X-mode will be reflected almost perfectly at the left hand cutoff (L-cutoff) layer and finally converted into EBW at UHR layer. In this experiment, the values of $\eta$ and $C_{max}$ are respectively about 0.8 and 0.3, so that the mode conversion should be possible. The total deposited power is about 0.5kW across the power step-down ($\Delta P_{in}$~5kW), and corresponds to about 10% absorption for injected power. This low absorption rate may be due to wave power loss to the outside of plasma through multiple reflection caused by poor wave directivity and low one-path absorption rate. The other is that the scale length of electron density profile is not sufficiently small in the edge, compared with the wavelength. Moreover, large density fluctuations near the edge might degrade the purity of wave polarization and the conversion rate to EBW.

**3.2 Perpendicular injection to target over-dense plasma**

In order to clarify and increase the mode conversion rate to EBW, we tried to superimpose the perpendicular injection of microwave power as X-mode in the course of a discharge produced by obliquely injected high power microwaves of 15kW. The target plasma for the additional perpendicular injection has already reached high density having a steep density gradient near the edge. In Fig.5(a), the time evolutions of $n_e^{2mm}$ and the electron kinetic energy ($E$) are compared for both cases with and without the additional perpendicular injection, where E is evaluated as

$$E = \int_V \frac{3}{2} n_e T_e(\rho) dV . \qquad (3)$$

The X-mode injection increase $n_e^{2mm}$ and $E$ by about 15% and 75% compared to the case with oblique injection only, respectively. The radial profiles of $n_e$ and $T_e$ at $t$=110 ms with and without the perpendicular injection, and $P_{abs}$ derived by eq.(1) are shown in Fig.5(b). The electron temperature is increased over the whole radial region by the perpendicular injection, where the $n_e$ profile remains unchanged by the injection. The additional perpendicular injection clearly contributes to clear electron heating. Although the cutoff layers are condensed near LCFS, the observed absorption of injection power widely takes place beyond these cutoff layers. The values of $\eta$ and $C_{max}$ in the target plasma are respectively about 0.43 and 0.76 at the moment of the perpendicular injection. The total deposited power is evaluated to be about 5kW across the power step-up ($P_{in}$~14kW), and corresponds to be fairly high value of about 35% absorption rate.

**4. Conclusion**

In the plasmas generated by the perpendicular injection microwaves with X-mode polarization at very low $B_t$ (<0.1T) on CHS, the central electron temperature reaches around 20eV and electron density reaches more than twice or 3 times of the O-mode cutoff density. The power is absorbed inside the over-dense region. The results are interpreted that the EBW converted from launched X-mode plays an essential role in this effective electron heating and over-dense plasma production. So far, the power absorption efficiency is still low around 10 to 20% for the incidence power. This may be caused by worse wave directivity and opaque polarization of incidence microwaves. Recently, we succeeded in raising the absorption efficiency by a factor of two by the use of the perpendicular injection of X-mode after the formation of relatively high density plasma having steep edge density gradient.

As a future research, the following issues are left for important and challenging targets.

1. Observation of parametric decay which will occur in X-B mode conversion process[11].
2. O-X-B conversion by launching obliquely O-mode.
3. Comparison of experimentally observed absorption rate with theoretical results.
4. Detailed power balance study including measurement of local radiation loss and neutral particle effect

Acknowledgements

This research was supported in part by the Granty-in-Aid for Scientific Research from JSPS, No. 15206107.

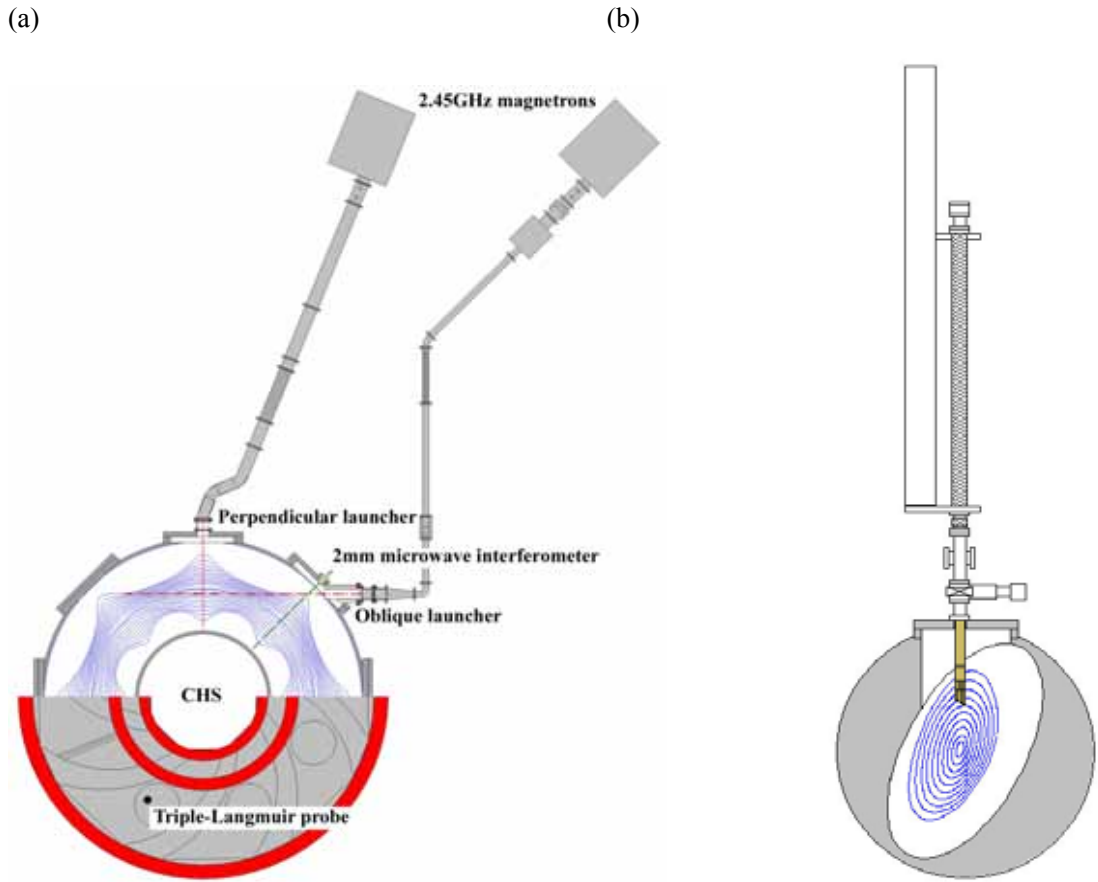

Fig.1 (a)Location of launching ports on CHS. (b)Arrangement of a triple-Langmuir probe on CHS.

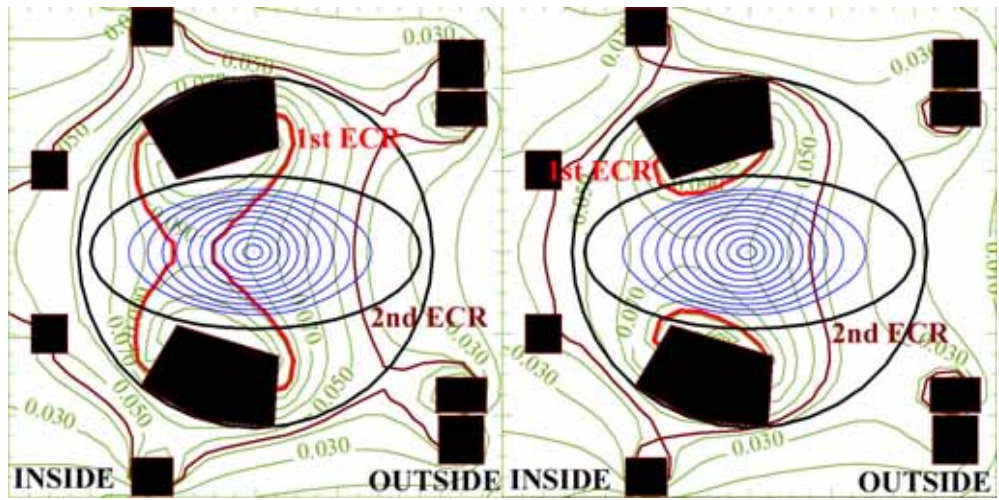

Fig.2 Contours of the magnetic field strength for the cases of $B_t$=0.0788T (left) and 0.0613T (right), together with the magnetic surface of the configuration of $R_{ax}$=97.4cm. Red, brown and blue lines are respectively fundamental ECR layer, 2nd ECR layer and magnetic surface.

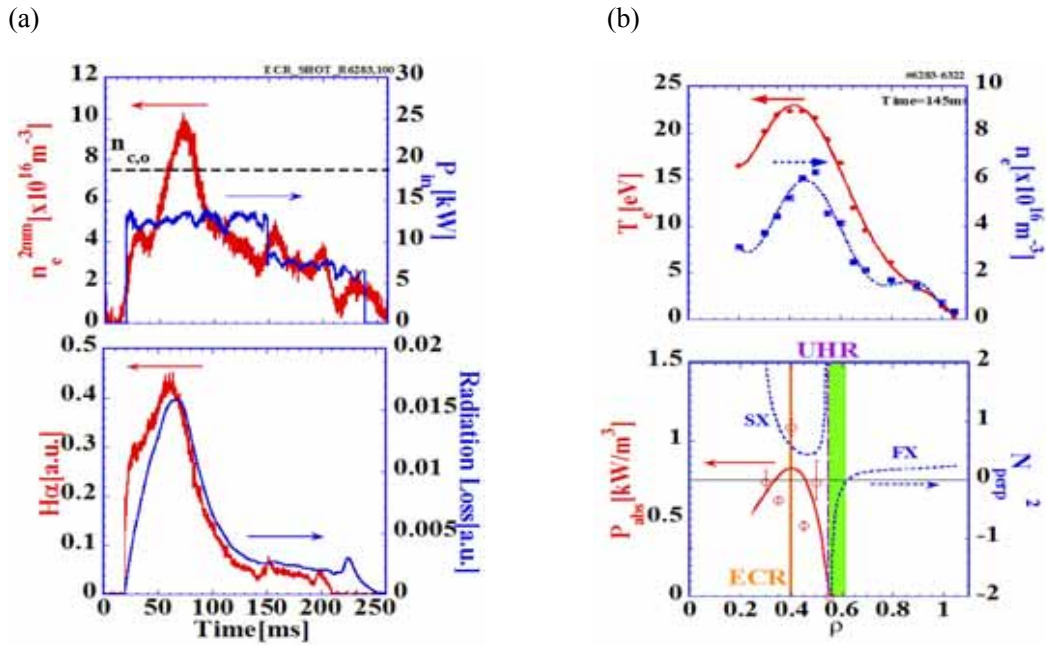

Fig.3 (a)Time evolutions of line averaged electron density ($n_e^{2mm}$), injected microwave power ($P_{in}$), Hα emission ($H_\alpha$) and total radiation loss. (b)Radial profiles of electron temperature ($T_e$), electron density ($n_e$), effective power absorption density ($P_{abs}$), and squared refractive index perpendicular to toroidal field ($N_{perp}^2$). Green zone is evanescent regions of fast X-mode.

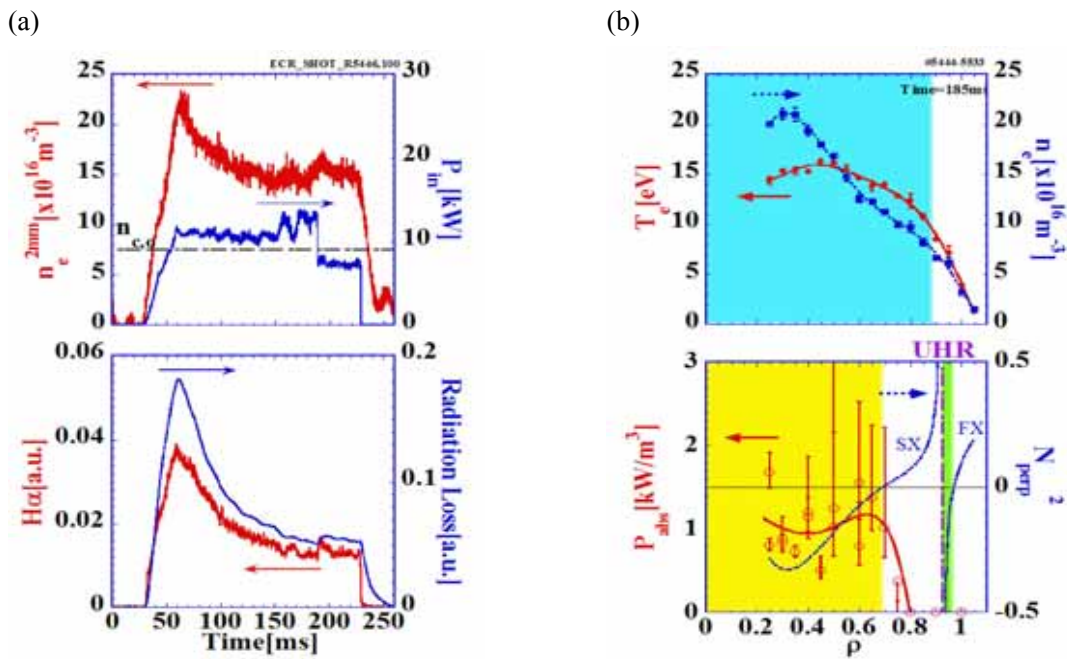

Fig.4 (a)Time evolutions of $n_e^{2mm}$, $P_{in}$, $H_\alpha$ and total radiation loss. (b)Radial profiles of $T_e$, $n_e$, $P_{abs}$, and $N_{perp}^2$. Blue, green and yellow zones are respectively evanescent regions of O-mode, fast X-mode and slow X-mode.

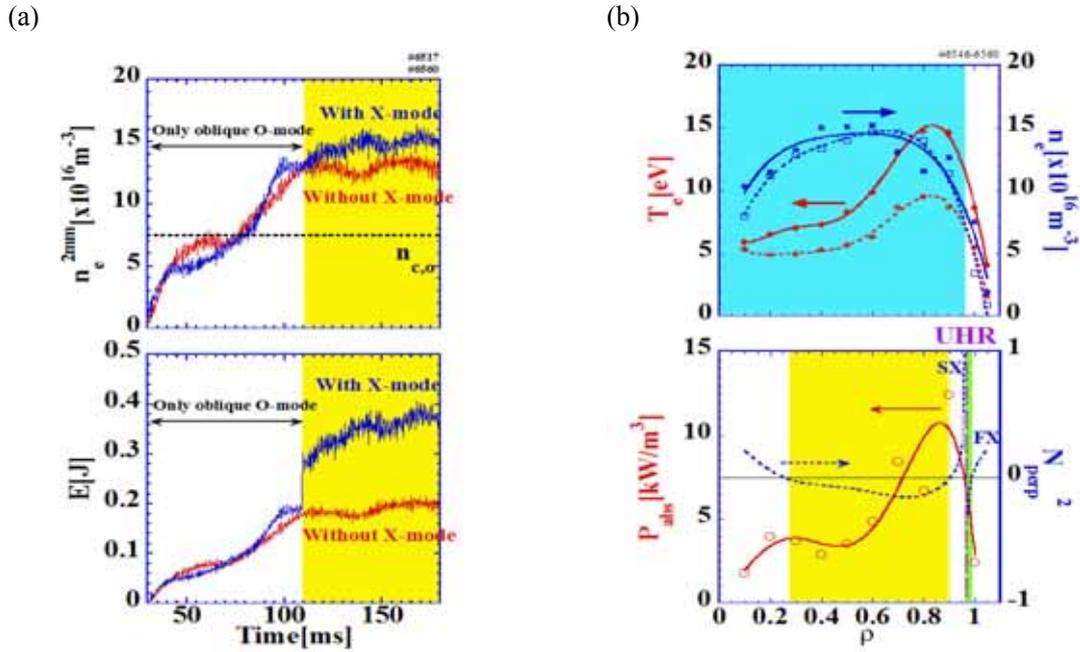

Fig.5 (a)Time evolutions of $n_e^{2mm}$ and electron kinetic energy ($E$) before launching X-mode and after. (b) Radial profiles of $T_e$, and $n_e$ with and without the additional perpendicular injection. The deposited power density of the perpendicular injection is shown as $P_{abs}$. Blue, green and yellow zones are respectively evanescent regions of O-mode, fast X-mode and slow X-mode.